# Searching for Superconductivity in the Z = 12.67 Family

O. P. Isikaku-Ironkwe[1, 2]
[1]The Center for Superconductivity Technologies (TCST)
Department of Physics,
Michael Okpara University of Agriculture, Umudike (MOUAU),
Umuahia, Abia State, Nigeria
and
[2]RTS Technologies, San Diego, CA 92122

## Abstract

The Z = 12.67 family of materials includes CaBeSi and NaAlSi which have been reported as non-superconducting and superconducting respectively. We show with an algorithm how to generate other members of this family. We discover through the material specific characterization dataset (MSCD) scheme and the symmetry conditions for superconductivity, why some are superconducting and others may not be superconducting.

## Introduction

The discovery of superconductivity in a simple binary compound magnesium diboride at 39K [1, 2] suggested that other materials may exist with similarly high transition temperatures, Tc. Borides and iso-structural and iso-electronic materials were searched [3 - 12] without achieving Tcs near magnesium diboride's. DFT and Eliasberg Theory computations [4,13] and superconductivity DFT theories [14, 15], powerful as they are in predicting electronic structure, have not revealed the true building blocks of material specificity in superconductivity [16, 17]. A material specific theory should be able to predict correctly, a priori, the existence of superconductivity in a material and estimate within reasonable error its Tc [17]. Studying superconductivity from a chemical viewpoint, we looked at correlations of superconductivity with Periodic Table properties such as electronegativity, valence electron count, atomic number and formula weight. These correlations [18] have guided our search for novel superconductors and material specificity of superconductivity and Tc [17]. We have made some original predictions of superconductors [19 - 26] in the broad $MgB_2$ family that await experimental confirmation or refutation. Two members of the Z=12.67 family have already been studied [27, 28] and give apparently conflicting results that are yet to be resolved. In this paper, we explore further the Z =12.67 family for superconductivity, estimating their Tcs using the material specific methodology [16] we have developed.

## Design and Search for Z =12.67, Ne= 2.667 Materials

Two well-studied materials, NaAlSi [27] and CaBeSi [28] belong to the family of materials with Z =12.667 and Ne=2.667. We can discover other members of this family using this simplified 5-steps algorithm, the Periodic Table and Tables 1, 2 and 3:

1. The sum of the valence electrons for 3-atom ternary material must be 8 or 16 for 6-atom ternary.
2. The sum of the atomic numbers that make up the valence electrons for 3-atom ternary must be 38 or 76 for 6-atom ternary. .
3. The formula is obtained by converting the atomic numbers in step 2 into the elements
4. The average electronegativity and formula weight are computed as shown in references [16]
5. The material so obtained must be thermodynamically stable which can be verified by DFT calculations or experimentally.

With this algorithm we obtain possible 3-atom members of the Z=12.67 family. For the 6-atom members obtained in Table 4 (Nos. 15 & 16), the algorithm for electronegativity, valence electron count, Ne, and atomic number, Z, is :

$$\frac{1}{6}(3G1 + 1G3 + 2G5) \qquad (1)$$

where G1, G3 and G5 represent the values of $\chi$, Ne, Z for the selected elements in Groups 1, 3 and 5 respectively of the Periodic Table partially produced in Table 3 for $\chi$ and Ne, using Pauling's electronegativity scale and Matthias valence electron convention first described in [16].

## Superconductivity, Tc and the Z =12.667 Family

We showed in [16] that superconductivity in a material may be estimated by:

$$0.75 < Ne/\sqrt{Z} < 1.02 \qquad (2)$$

The parameters of MSCD are computed as shown in [16] and [26]. From the MSCDs in Table 4, we see that the Z =12.667 family meets this criteria and is likely to show superconductivity with maximum transition temperature, Tc given [16] by:

$$T_c = \mathcal{X} \frac{Ne}{\sqrt{Z}} K_o \qquad (3)$$

Ko is a parameter related to Fw/z by Ko = n(Fw/Z) where n can be estimated empirically. For $MgB_2$ we found n =3.65. For NaAlSi, with Tc =7K, n = 1.083. The estimated Tcs for members of the Z =12.667 are displayed in Table 4. We have assumed n =1.083. For $K_3BN_2$ we have computed Tc with three values of n, indicating that the Tc could be as low as 17.3K or as high as 58.4K with a mean around 37.9K

## Analysis & Discussion

A unique feature of the Z=12.67 family is that they all have the same Ne and Z and show very close Fw. MSCD analysis shows that NaAlSi and CaBeSi have the same Z and Ne and almost the same $\mathcal{X}$. From the symmetry rules discovered in [16] their Tcs should be very close with CaBeSi having a slightly higher Tc. However, computation in reference [28] puts the Tc of CaBeSi <1K. The experimentally obtained value for the Tc of NaAlSi is 7K. Thus we compute the Tc of CaBeSi to be 7.1K. CaBeSi's Tc need therefore to be verified independently. Using the computed n from NaAlSi of 1.083 we have estimated the Tc of other members of the Z =12.67 family which are displayed in Table 4. Of particular interest is the material $K_3BN_2$ which is a 6-atom ternary with electronegativity and valence electron count the same as $MgB_2$. From symmetry rules in [16] their Tc should be proportional to their Zs. Using n =3.65 and equation (3) we estimate an upper bound Tc of 58.4K for $K_3BN_2$. Taking the mean of 1.083 and 3.65, the Tc comes to 37.9K. Checking the Tc experimentally will be an indirect check on the symmetry rules and Tc estimation equation above. The other members of the Z=12.67 family need too to be experimentally verified as superconducting at the estimated Tcs. The Z=12.67 and the Z=4.67 [26] seem to present materials that may combine to yield an intermediate Z say 7.33. This possibility needs further investigation in a future study.

## Conclusion

We showed how to generate new members of the Z =12.67 family with Ne =2.67. Two ternary types exist: the 3-atom and 6-atom types. The Z = 12.67 family do meet the condition for superconductivity that is $0.75 < Ne/\sqrt{Z} < 1.02$. Electronegativities vary within this family reaching 1.733 in $K_3BN_2$ as in $MgB_2$. Our estimated Tc for CaBeSi is much higher than obtained elsewhere [28] and requires further experimental investigation. Some Z=12.67 members with estimated higher Tc such as $K_3BN_2$, NaCaN, KAlC, KBSi and $LiBGe_{0.89}Si_{0.11}$ require further study by DFT methods and also experimentally to verify their existence and the estimated Tcs.

## Acknowledgements

The author acknowledges useful and inspiring discussions at various times in the course of this study with J.R. O'Brien at Quantum Design, A.O.E. Animalu at University of Nigeria, M. J. Schaffer formerly at General Atomics, San Diego and M.B. Maple and J.E. Hirsch at UC San Diego.

## Tables

| | | |
|---|---|---|
| 1 | 1 | 6 |
| 1 | 2 | 5 |
| 1 | 3 | 4 |
| 2 | 2 | 4 |
| 2 | 3 | 3 |

**Table1:** Combinations of valence electrons that sum to 8 for a 3-atom ternary/binary system. The valence electrons represent the group in the Periodic Table. For 6-atom compound the algorithm is given in (1).

| Group | Possible atomic numbers, Z | | | |
|---|---|---|---|---|
| 1 | 1 | 3 | 11 | 19 |
| 2 | 4 | 12 | 20 | - |
| 3 | 5 | 13 | 31 | - |
| 4 | 6 | 14 | 22 | 32 |
| 5 | 7 | 15 | 23 | 33 |
| 6 | 8 | 16 | 34 | - |

**Table 2:** Possible atomic numbers to combine from six groups to get 38 for 3-atom system with sum of valence electron of 8.

| Ne = 1 | | Ne = 2 | | Ne = 3 | | Ne = 4 | | Ne = 5 | | Ne = 6 | |
|---|---|---|---|---|---|---|---|---|---|---|---|
| E | $\mathcal{X}$ | E | $\mathcal{X}$ | E | $\mathcal{X}$ | E | $\mathcal{X}$ | E | $\mathcal{X}$ | E | $\mathcal{X}$ |
| H | 2.1 | Be | 1.5 | Sc | 1.3 | Ti | 1.5 | V | 1.6 | Cr | 1.6 |
| Li | 1.0 | Mg | 1.2 | Y | 1.2 | Zr | 1.4 | Nb | 1.6 | Mo | 1.8 |
| K | 0.8 | Ca | 1.0 | La | 1.1 | Hf | 1.3 | Ta | 1.5 | W | 1.7 |
| Rb | 0.8 | Sr | 1.0 | B | 2.0 | C | 2.5 | N | 3.0 | O | 3.5 |
| Cs | 0.7 | Ba | 0.9 | Al | 1.5 | Si | 1.8 | P | 2.1 | S | 2.5 |
| Cu | 1.9 | Zn | 1.6 | Ga | 1.6 | Ge | 1.8 | As | 2.0 | Se | 2.4 |
| Ag | 1.9 | Cd | 1.7 | In | 1.7 | Sn | 1.8 | Sb | 1.9 | Te | 2.1 |
| | | Hg | 1.9 | Tl | 1.8 | Pb | 1.8 | Bi | 1.9 | | |

| Ne = 7 | | Ne = 8 | | Ne = 9 | | Ne = 10 | | Nomenclature |
|---|---|---|---|---|---|---|---|---|
| E | $\mathcal{X}$ | E | $\mathcal{X}$ | E | $\mathcal{X}$ | E | $\mathcal{X}$ | |
| Mn | 1.5 | Fe | 1.8 | Co | 1.8 | Ni | 1.8 | $\mathcal{X}$ = Electronegativity |
| Tc | 1.9 | Ru | 2.2 | Rh | 2.2 | Pd | 2.2 | E = Element |
| Re | 1.9 | Os | 2.2 | Ir | 2.2 | Pt | 2.2 | Ne = Valence Electron Count |

| Ne = 3 |
|---|
| E | Ce | Pr | Nd | Pm | Eu | Tb | Yb | Lu | Sm | Gd | Ho | Er | Tm | Th | Pu | U |
|---|---|---|---|---|---|---|---|---|---|---|---|---|---|---|---|---|
| $\mathcal{X}$ | 1.1 | 1.1 | 1.1 | 1.1 | 1.1 | 1.1 | 1.1 | 1.1 | 1.2 | 1.2 | 1.2 | 1.2 | 1.3 | 1.3 | 1.3 | 1.4 |

| F | 4.0 |
|---|---|
| Cl | 3.0 |
| Br | 2.8 |
| I | 2.5 |

Table 3: Electronegativity Scale and Valence Electron Count Convention from Paul and Matthias(adapted from [16])

| | Material | $\mathcal{X}$ | Ne | Z | Ne/$\sqrt{Z}$ | Fw | Fw/Z | n | $K_0$ | Tc |
|---|---|---|---|---|---|---|---|---|---|---|
| 1 | CaBeSi | 1.433 | 2.667 | 12.667 | 0.7493 | 77.18 | 6.09 | 1.083 | | 7.1 |
| 2 | Na$_2$S | 1.433 | 2.667 | 12.667 | 0.7493 | 78.05 | 6.16 | 1.083 | | 7.2 |
| 3 | LiKS | 1.433 | 2.667 | 12.667 | 0.7493 | 78.11 | 6.17 | 1.083 | | 7.2 |
| 4 | **NaAlSi** | **1.4** | **2.667** | **12.667** | **0.7493** | **78.06** | **6.16** | **1.083** | **6.673** | **7.0** |
| 5 | Mg$_2$Si | 1.4 | 2.667 | 12.667 | 0.7493 | 76.71 | 6.06 | 1.083 | | 6.9 |
| 6 | NaMgP | 1.4 | 2.667 | 12.667 | 0.7493 | 78.27 | 6.18 | 1.083 | | 7.0 |
| 7 | KAlC | 1.6 | 2.667 | 12.667 | 0.7493 | 78.09 | 6.16 | 1.083 | | 8.0 |
| 8 | NaCaN | 1.633 | 2.667 | 12.667 | 0.7493 | 77.08 | 6.09 | 1.083 | | 8.1 |
| 9 | LiBGe$_{0.89}$Si$_{0.11}$ | 1.6 | 2.667 | 12.667 | 0.7493 | 85.46 | 6.75 | 1.083 | | 8.8 |
| 10 | KBSi | 1.533 | 2.667 | 12.667 | 0.7493 | 78.0 | 6.16 | 1.083 | | 7.7 |
| 11 | KBeP | 1.467 | 2.667 | 12.667 | 0.7493 | 79.09 | 6.24 | 1.083 | | 7.4 |
| 12 | MgAl$_2$ | 1.4 | 2.667 | 12.667 | 0.7493 | 78.27 | 6.18 | 1.083 | | 7.0 |
| 13 | LiAlTi | 1.333 | 2.667 | 12.667 | 0.7493 | 81.8 | 6.46 | 1.083 | | 7.0 |
| 14 | LiCaP | 1.367 | 2.667 | 12.667 | 0.7493 | 77.99 | 6.16 | 1.083 | | 6.8 |
| 15 | Na$_3$AlP$_2$ | 1.4 | 2.667 | 12.667 | 0.7493 | 157.89 | 12.465 | | | 14.2 |
| 16 | **K$_3$BN$_2$** | **1.7333** | **2.667** | **12.667** | **0.7493** | **156.13** | **12.326** | **1.083** <br> **2.37** <br> **3.65** | | **17.34** <br> **37.9** <br> **58.43** |

**Table 4:** 16 MSCDs of Z =12.667 Materials. $\mathcal{X}$ represents the electronegativity, Ne the valence electron count, Z atomic number and Fw the formula weight. MSCD computations were described in [16]. Note that Ne and Z remain constant while $\mathcal{X}$ changes for members of a family. Since 0.75<Ne/$\sqrt{Z}$ <1.02, they may be superconducting and Tc can be estimated as : **T$_c$ =** $\mathcal{X} \dfrac{Ne}{\sqrt{Z}}$ K$_o$ where Ko =n(Fw/Z) and n = 1.083 derived experimentally from NaAlSi with Tc=7K.